\documentclass[prl,nofootinbib,twocolumn,superscriptaddress]{revtex4}
\usepackage{epsfig}
\usepackage{axodraw}

\newcommand{\la}{{\lambda}}

\newcommand{\be}{\begin{equation}}
\newcommand{\ee}{\end{equation}}
\newcommand{\beq}{\begin{equation}}
\newcommand{\eeq}{\end{equation}}
\newcommand{\bea}{\begin{eqnarray}}
\newcommand{\eea}{\end{eqnarray}}
\newcommand{\br}{\begin{eqnarray}}
\newcommand{\er}{\end{eqnarray}}
\newcommand{\ba}{\begin{array}}
\newcommand{\ea}{\end{array}}
\newcommand{\bi}{\begin{itemize}}
\newcommand{\ei}{\end{itemize}}
\newcommand{\bn}{\begin{enumerate}}
\newcommand{\en}{\end{enumerate}}
\newcommand{\bc}{\begin{center}}
\newcommand{\ec}{\end{center}}

\newcommand{\bT}{\bar{T}}
\newcommand{\tl}{\tilde{L}}
\newcommand{\ta}{\tilde{a}}
\newcommand{\tH}{\tilde{H}}

\newcommand{\no}{\nonumber}
\newcommand{\nn}{\nonumber}

\newcommand{\eps}{\epsilon}
\newcommand{\eq}[1]{eq.(\ref{#1})}

\newcommand{\rfn}[1]{(\ref{#1})}

\newcommand{\gsim}{\lower.7ex\hbox{$\;\stackrel{\textstyle>}{\sim}\;$}}
\newcommand{\lsim}{\lower.7ex\hbox{$\;\stackrel{\textstyle<}{\sim}\;$}}

\def\mysection#1{\noindent {\bf #1} }

\begin{document}
%\draft

%\preprint

\title{\bf 
Leptogenesis in the minimal 
supersymmetric triplet seesaw model}

\author{Giancarlo D'Ambrosio}
\affiliation{ INFN, Sezione di Napoli and Dipartimento di 
Scienze Fisiche,  Universit\`a di Napoli, I-80126 Naples, Italy}
\author{Thomas Hambye}
\affiliation{Theoretical Physics Subdept, University of Oxford, 
1 Keble Road, Oxford OX13NP, UK}
\author{Andi Hektor}
\author{Martti Raidal}
\affiliation{National Institute of Chemical Physics and Biophysics, 
Ravala 10, Tallinn 10143,  Estonia} 
\author{Anna Rossi}
\affiliation{Dipartimento di Fisica ``G.~Galilei'', Universit\`a di Padova 
and INFN, Sezione di Padova, I-35131 Padua, Italy}

\date{\today}
%\pacs{}

\vspace*{0.in}

\begin{abstract}
\noindent
In the  supersymmetric triplet (type-II) seesaw model, 
in which a single 
$SU(2)_L$-triplet couples to leptons, 
the high-energy neutrino flavour structure 
can be directly determined from the low-energy neutrino data.
We show that  even with such a minimal triplet content, 
leptogenesis 
can  be naturally accommodated thanks to the   
resonant interference between superpotential and
soft supersymmetry breaking terms.

\end{abstract}

\maketitle

\mysection{{\bf 1.}~Introduction.}~Non-zero 
neutrino masses and mix\-ing angles 
provide the only convincing  evidence~\cite{nuexp} of physics 
beyond the standard model we have today. 
The seesaw mechanism can be regarded as a paradigm to 
understand these neutrino properties. 
According to this mechanism at the decoupling of some heavy states with 
mass larger than the electroweak scale, and 
coupled to $SU(2)_L$ lepton and Higgs doublets, the $d=5$ 
operator 
$~\frac{\cal K}{M} L L  H H$
is generated, where $M$ is a typical scale at which the  
lepton number is violated.
Upon breaking of $SU(2)_L\times U(1)_Y$, 
by $\langle H\rangle = v$,
Majorana masses for the neutrinos 
are induced, $m_{\nu}  = \frac{v^2}{M} {\cal K}$.   
In the basis in which  
the charged-lepton mass  matrix  is diagonal,  
all the low-energy  flavour structure  is contained 
in the  matrix ${\cal K}$, i.e. in the neutrino mass matrix.
For neutrino masses in the range 
$\sim (10^{-1} - 1)~{\rm eV}$,  the overall scale $M$ can be inferred, 
${\cal K}^{-1} M \sim 10^{14} ~{\rm GeV}$.

The most popular realizations 
of the seesaw  mechanism are the singlet seesaw \cite{seesaw} (or type-I) 
in which three singlet fermions $N$ are exchanged, 
and the triplet \cite{valle,tseesaw} seesaw \cite{tseesaw} (or 
type-II) in which a 
$SU(2)_L$ triplet scalar $T$ with non-zero hypercharge is 
exchanged\footnote{In the supersymmetric scenario,
which we assume in this paper from the very beginning, 
the anomaly cancellation requires minimally a pair of triplets
$T,\bar T$ with opposite $U(1)_Y$ quantum numbers. However, only one
triplet $T$ couples to leptons.}.
The two realizations differ in an important aspect. Namely,  
the singlet seesaw has two sources of flavour structure, 
the Yukawa coupling matrix $Y_N$ of the singlets $N$ with the 
$SU(2)_L$ leptons, and the mass matrix $M_N$ of the states $N$, 
which combine to give ${\cal K}/M  =  Y_N^T M^{-1}_N Y_N$.  
The triplet seesaw scenario has only one source of flavour, {\it i.e.} 
the symmetric Yukawa matrix $Y_T$ of the triplet $T$ couplings with 
the leptons, and  so $ {\cal K}/M = \la_2 Y_T/M_T$, where 
 $M_T$ is the triplet mass and  $\la_2$ is a dimensionless 
(`unflavoured') Higgs coupling. This implies that 
in the type-I seesaw case the high-energy quantities, $Y_N$ and $M_N$ 
which contain   18 parameters\footnote{
Predictive seesaw models,  with two $N$'s
are also viable~\cite{fgy}.} 
cannot be univocally related 
to the 9 low-energy neutrino parameters, 
encoded in the symmetric neutrino mass matrix $m_\nu$.
Instead in the type-II case 
the high-energy quantity $Y_T$, which 
contains 9 parameters 
can be matched to the  flavour structure of the matrix $m_\nu$
\cite{tseesaw,anna}.
What is left undetermined is the overall mass scale $M_T/\la_2$.

As both the seesaw scenarios involve a high-energy scale where  
the lepton number $L$ is not conserved, either $M_N$ or $M_T$,  
they also offer a possibility to generate 
the observed baryon asymmetry of the Universe via out-of-equilibrium decays 
of these heavy states.
In the type-I seesaw model leptogenesis~\cite{fy} occurs naturally 
(both in its supersymmetric as well as non-supersymmetric versions) 
since there are three heavy singlet neutrinos. 
However, the situation is more involved in the type-II seesaw case.
The minimal triplet seesaw with one-pair $T,\bar{T}$ 
(or with one triplet $T$ in the  non-supersymmetric case)
does contain sources of $L$ and CP violation at high scale, 
but it cannot effectively induce leptogenesis \cite{ms,hms}.
Successful leptogenesis has been achieved by adding one more 
vector-like  $T^{'}, \bar{T}^{'}$ pair  \cite{ms,hms}, 
and therefore one more  Yukawa matrix $Y_{T^{'}}.$ 
In such a case 9 more flavour parameters will  be added,
and the direct connection between the high-energy flavour structure and 
the low-energy one is lost. Alternatively, one can 
consider the hybrid type-I plus type-II model
 (motivated by left-right and SO(10) models) where a single triplet 
couples to leptons \cite{utpal,recent,otherhybrid}.
In this case, successful leptogenesis can be obtained in agreement 
with neutrino data
from the interplay of type-I and type-II contributions \cite{recent},
though the above flavour connection is also lost.

In this paper we show that leptogenesis turns out to be  
in fact possible in the minimal 
supersymmetric triplet seesaw scenario with 
{\it only one} pair of triplets $T, \bT$.
Nonvanishing interfering amplitudes 
emerge due to soft supersymmetry 
breaking mass parameters (not considered in \cite{hms}), and resonant
``soft'' leptogenesis~\cite{nir,dgr} occurs\footnote{
Other scenarios of (non-resonant)  leptogenesis induced by 
soft supersymmetry breaking terms can be found in \cite{others,GKNR}. 
See also \cite{boub}.}. 
As a result the observed $B-L$ asymmetry
can be generated due to finite temperature effects~\cite{gnrrs}.

This paper is organized as follows. In Section 2 we introduce 
the relevant superpotential and Lagrangian with the soft-breaking terms 
to describe the minimal triplet seesaw framework. 
In Section 3 the lepton asymmetry is derived including the 
thermal corrections,  
and hence the baryon asymmetry is obtained. 
We discuss our results in Section 4, 
give a few general remarks in Section 5 and  
conclude in Section 6.

\vspace{0.3cm}

\mysection{{\bf 2.}~The triplet-seesaw scenario.}~The supersymmetric 
tri\-plet seesaw model 
is described by the following superpotential
\be
W\!= \!
\frac{1}{\sqrt{2}}(Y^{ij}_T L_i T L_j + 
\lambda_1 H_1 T H_1 + \lambda_2 H_2\bar T H_2)+ M_T T \bar T , 
\label{W}
\ee
where $i,j\!\!= \!\!e, \mu ,\tau$ 
are flavour indices, $T\!\!=\!\!(T^0, T^{+},T^{++})$, 
$\bar T\!=\!(\bT^{0}, \bT^{-}, \bT^{--})$ are in a vector-like 
$SU(2)_L \times U(1)_Y$ representation, $T\sim (3,1)$ and $\bT \sim(3,-1)$, 
$L_i$  are the lepton  $SU(2)_L$ doublets, 
and  $H_{1} (H_2)$ is the Higgs doublet with hypercharge -1/2 (1/2).
In eq.~\rfn{W}, the symmetric matrix $Y_T$ 
contains  6 real parameters plus 3 phases, 
while the parameters $\la_2$ 
and $M_T$ can be taken to be  real, and  $\la_1$  is in general complex.
We  observe that if, for example, 
we assign
%\footnote{This is the $L$-assignment we assume 
%and in the following the definition of the lepton asymmetry 
%will be accordingly given. Of course, the result does not depend 
%on such an assignment.}
the lepton number $L=0$ to the 
triplets $T, \bT$, the $Y_T$-couplings explicitly break $L$, while 
the $\lambda_1, \lambda_2$ couplings conserve $L$.  
By integrating out the triplet states at the scale $M_T$, 
the $d=5$  $L$-violating  operator is generated which after the
electroweak symmetry breaking induces the neutrino mass matrix
\be
m_\nu^{ij}=
Y_T^{ij} \lambda_2 \frac{v_2^2}{M_T} .
\label{mnu}
\ee
This clearly shows that the flavour structure of neutrino mass matrix 
is directly linked to that of the matrix $Y_T$.  

In the supersymmetric limit the states $T$ and $\bT$ are degenerate 
with mass $M_T$. The necessary mass splitting  and CP-violation 
to induce resonant 
leptogenesis is provided by the 
soft supersymmetry breaking (SSB) terms. The relevant ones are:
\bea
\label{Lsoft}
-{\cal L}_{soft} & = &
\left[ \frac{1}{\sqrt{2}}( A^{ij}_{T }{\tilde L}_i T {\tilde L}_j
+ A_1 H_1TH_1 + A_2 H_2\bar T H_2 ) 
\right. \no \\
&& \left. 
+B M_T T\bar T + {\rm h.c.}\right] 
+{\tilde m}^2_{T} T^\dagger T +  {\tilde m}^2_{\bar T} 
{\bar T}^\dagger\bar T , 
\eea
where a  standard notation has been adopted. 
In eq.~\rfn{Lsoft},  the matrix $A_T$ 
has 6 real parameters plus 3 phases, and  
the mass parameters $A_1, A_2$ and $(B M_T)$ are  in general complex. 
In the following, 
we set to zero ${\rm arg}(B M_T)$  without loss of generality. 
The  $B$-term in eq.~\rfn{Lsoft} removes the mass degeneracy, 
$M^2_\pm=M^2_T \pm B M_T$ (where 
the SSB terms ${\tilde m}^2_{T (\bT)}\!\ll\! M^2_T$ have been neglected), 
between the mass eigenstates $T_\pm$ which 
are given by\footnote{We have adopted the 
representation for $T$ and $\bT$ as given in \cite{anna}. 
Therefore, in the r.h.s. of eq.~(\ref{tpm}) the term  $\bar{T}^*$ 
is understood as $i\sigma_2 \bar{T}^* i\sigma_2$, where 
$\sigma_2$ is the complex Pauli matrix.} 
\beq\label{tpm}
T_\pm= \frac{1}{\sqrt{2}}\left( T\pm \bar{T}^* \right).
\eeq
In the $T_\pm$-basis the relevant Lagrangian interactions 
read as:
\bea
\label{mb}
-{\cal L}& = & \frac{1}{\sqrt2} \left( 
Y^{ij}_{T} L_i \frac{T_{+} +T_{-} }{\sqrt2} L_j
+{\lambda_{1}} \tilde{H}_1 
\frac{T_{+} +T_{-} }{\sqrt2}\tilde{H}_1   \right.  \no \\ 
&& \left. + {\lambda_2} \tilde{H}_2 
\frac{T^*_{+} - T^*_{-} }{\sqrt2} \tilde{H}_2   +
{M_T Y^{ij}_{T}} \tilde{L}_i  \frac{T_{+} -T_{-} }
{\sqrt2} \tilde{L}_j \right. \no \\
&& \left. +
{M_T \la_1} H_1   \frac{T_{+} -T_{-} }
{\sqrt2} H_1 +
{M_T \la_2} H_2   \frac{T^*_{+} + T^*_{-} }
{\sqrt2} H_2 \right.\no \\
&& \left. +{A^{ij}_{T}} \tilde{L}_i  \frac{T_{+} + T_{-} }
{\sqrt2} \tilde{L}_j + {A_1} H_1   \frac{T_{+} +T_{-} }
{\sqrt2}H_1 \right.\no \\ 
&&\left. + {A_2} H_2  \frac{T^*_{+} -T^*_{-} }
{\sqrt2}  H_2 \right)
 + {\rm h.c.} 
\label{lag}
\eea
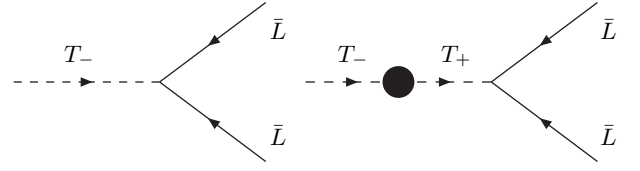
\begin{figure}[t]
\begin{center}
\begin{picture}(220,60)(0,0)
\ArrowLine(95,0)(55,30)
\ArrowLine(95,60)(55,30)
\DashArrowLine(0,30)(55,30){3}
\Text(25,40)[]{\small $ T_-$}
\Text(100,50)[]{\small $\bar L$}
\Text(100,10)[]{\small $\bar L$}
%\Text(110,40)[]{\small $f_-$}
%%%%%%%%%%
\ArrowLine(220,0)(180,30)
\ArrowLine(220,60)(180,30)
\DashArrowLine(110,30)(145,30){3}
\Vertex(145,30){6}
\DashArrowLine(145,30)(180,30){3}
\Text(130,40)[]{\small $ T_-$ }
\Text(169,40)[]{\small $ T_+$ }
\Text(225,50)[]{\small $\bar L$}
\Text(225,10)[]{\small $\bar L$}
%\Text(300,40)[]{\small $f_+$}
%%%%%%%%%%%%%%%%%%%%%%
\end{picture}
\end{center}
\caption{\it 
Interfering $T_-$ decay amplitudes for the fermionic final states. 
Analogous diagrams exist for bosonic final states. 
The two-point function $\Pi_{+-}$ is denoted by a blob.
}  
\label{fig1}
\end{figure}

\vspace{0.3cm}

\mysection{{\bf 3.} The lepton asymmetry.}
To take into account the resonant phenomenon of CP-violation 
due to the mixing of (almost) degenerate states, as $T_\pm$ are,  
we use an effective field-theory approach  of resummed propagators 
for unstable (mass eigenstate) particles, as described in ref.~\cite{p}.
The decay amplitude ${\widehat S}_-^f$ 
of the unstable external state $T_-$ into a final state
$\bar f$ is described by a superposition of amplitudes $S_\pm^f$ 
with stable external states. Adding 
the contributions shown in Fig.~\ref{fig1}, we obtain
\beq
\widehat{S}_-^f(T_-  \rightarrow \bar f ) = S_-^f - S_{+}^f 
\frac{i\Pi _{+-}}{M_-^2 -M_+^2 + i\Pi _{++}} ,
\label{ampl}
\eeq
where $\Pi _{\alpha \beta} (p^2)$ are the absorptive parts of the
two-point functions for $\alpha,\beta =+,-.$ 
The amplitude for the decay into the conjugate final state is
obtained by the replacements $S_\pm\to S_\pm^*,$ $\Pi_{+-}\to\Pi_{+-}^*.$ 
From the amplitudes (\ref{ampl}) the corresponding widths can be 
obtained. 
The relevant $L$ violating decays are $T_\pm \to \bar{L}\bar{L} 
, \tl^* \tl^*$ with $\Delta L = -2$ and 
$T^*_\pm \to {L} {L},  \tl \tl$ with $\Delta L = 2$. 
The lepton asymmetry is obtained 
by calculating the $\Delta L$ produced on average each 
time  any component of $T_{\pm}$ or $T_{\pm}^*$ decays. 
Therefore, for $T_-$ and $T_-^*$ 
(and similarly for $T_+$ and $T_+^*$, by
interchanging $+$ and $-$) this is given by the following
CP asymmetry ($f=L_i L_j,~\tilde L_i \tilde L_j$): 
\bea
\eps_-&=&
\Delta L~ \frac{
\sum_f \left[\Gamma({T_-^*} \rightarrow f)-
\Gamma(T_- \rightarrow {\bar f})\right]
}{
\Gamma_{T_-^*}+\Gamma_{T_-}
} \no \nn
\\ &= &  
-\frac{1}{8 \pi}\frac{1}{M_T \Gamma}
\frac{ \left( M_-^2 - M_+^2 \right) \,
\sum_f {\rm Im}\left( S_-^{f*} S_+^f \Pi_{+-}\right) c_f}
{ \left( M_-^2 - M_+^2 \right)^2 + \Pi_{++}^2} ~, \no \\
&&
\label{epsg2}
\eea
where  $\Delta L =2$ and the amplitudes $S^f$ can be read off from \eq{lag}
as follows: $S=2 h$ ($h$ is the corresponding coupling in \rfn{lag}) 
for the scalar final states and $S=2 M_T h$ for the fermionic final states.
Notice that 
the denominator involves not only the decays into  leptons and sleptons, 
as the numerator does, 
but the 
total decay 
width $\Gamma_{T_-^*}=\Gamma_{T_-}\equiv \Gamma$ of any member of the 
$T_-$ and $T_-^*$ triplets with
\be
\label{Gtot}
\Gamma =  
\sum_A \Gamma^A_{0}~ c_{A} ~,
\ee
where $A=(a, \tilde{a})$ with 
$a = (L, \tH_1, \tH_2), \ta = (\tl, H_1, H_2)$, and 
the various partial decay widths into a pair of $a$ or 
$\ta$  
at zero temperature are given by
\bea
\label{gpart}
\Gamma^L_{0} &=& \Gamma^{\tilde{L}}_{0} = 
\frac{M_T}{32\pi} {\rm Tr}(Y_TY_T^\dagger) ~, \\
\label{gpart1}
\Gamma^{\tH_i}_{0} &=& \Gamma^{H_i}_{0} =
\frac{M_T}{32\pi} |\la_i|^2 
~, ~~~i=1,2 .
\eea
In eqs.~\rfn{epsg2} and (\ref{Gtot}) the
$c$ factors
are the contributions to the phase-space factors of the final states 
due to  finite-temperature effects:  
\bea
c_{a}&=&(1-2 x_a)\sqrt{1-4 x_a}
\left[ 1-n_F(E_a)\right]^2  
\label{cfeq},\\
c_{\tilde a}&=&\sqrt{1-4 x_{\tilde a}}
 \left[ 1+n_B(E_{\tilde a})\right]^2 .
\label{cbeq} 
\eea
They reduce to 1 for $T=0$.
Here we have defined:
\bea
&E_{a (\ta)}=\frac{m_T(T)}{2} ,~~~
x_{a (\ta)}\equiv \frac{m_{a (\ta)}^2(T)}{m_T^2(T)}&\\
&n_F(E)=\frac{1}{e^{E_a/T}-1},~~~n_B(E)=\frac{1}{e^{E_{\ta}/T}+1},&
\eea
where the thermal masses read as (see {\it e.g.} \cite{denis})
\bea
m_T^2(T)&=& 
M_T^2+\left(g_2^2+\frac{1}{2}g_Y^2 \right) T^2,\nn\\
m_{\tilde L}^2(T)&=&2 m_L^2(T)= m^2_{H_1}(T)=
\left( \frac{3}{8}g_2^2+\frac{1}{8}g_Y^2\right)T^2, \nn\\
m_{H_2}^2(T)&=&
\left( \frac{3}{8}g_2^2+\frac{1}{8}g_Y^2 + \frac{3}{4} y_t^2\right)T^2 .
\label{Tmass}
\eea
Here $g_2$ and $g_Y$ are the $SU(2)_L$ and $U(1)_Y$ gauge couplings, 
respectively,  and $y_t$ is the top Yukawa coupling 
(to be renormalized at the appropriate high-energy scale).

In eq.~\rfn{epsg2} and 
in the following  we keep only the lowest-order 
contribution in the soft terms, {\it i.e.} we use the approximation 
$\Gamma_{T_-}=\Gamma_{T_+}=\Gamma$ and $M_+=M_-=M_T$,
 whenever it concerns only higher-order corrections.
Regarding  the factors 
$\Pi_{\pm \pm}$ and $\Pi_{+-}$  in eq.~(\ref{epsg2}), 
the diagrammatic computations give:
\bea
&&\Pi_{\pm\pm}=
M_T \left[ \Gamma^L_{0} (R_L + R_{\tilde{L}}) +
\sum_i \Gamma^{\tH_i}_{0} (R_{\tilde{H}_i} + R_{H_i}) \right] \no \\
&& \label{G} \\
&&\Pi_{+-}=\Pi_{-+}^*=i\frac{M_T}{16\pi} {\rm Im}
\left[{\rm Tr}(A_T Y_T^\dagger)R_{\tilde L}(T) \right.\no \\
&& \left. + 
(A_1 \lambda^*_1) R_{H_1}(T)+
(\lambda_2 A^*_2) R_{H_2}(T)\right] ,
\label{P}
\eea
where $R_{\tilde{a} }$ and $R_a$ ($\tilde{a} = \tilde{L}, H_1, H_2$, ~
${a} =L, \tilde{H}_1, \tilde{H}_2$) 
are 
the finite temperature 
 stimulated emission and Pauli-blocking  factors for
the  bosons and fermions  
in the loops, respectively \cite{gnrrs}:
\be
R_{\tilde a}(T) = \sqrt{1-4 x_{\tilde a}}
\left[
1+ 2n_B(E_{\tilde a})+2 n_B^2(E_{\tilde a})\right] ,
\label{R}
\ee
while for the (cumbersome) 
expression of $R_a$ we refer the reader to 
eq.~D1 in \cite{gnrrs}.  
We recall that the $R$ factors are different from the $c$ factors 
in eqs.~(\ref{cfeq}) and (\ref{cbeq}),  
as explained in \cite{gnrrs}. Note also 
that only bosonic loops contribute to $\Pi_{+-}$ and therefore
the CP-asymmetries are enhanced by the stimulated emission factors in 
$R_a$.
As for the  $L$-asymmetry $\eps_+$   from 
the  $T_+$  decays, we have that $\eps_+ = \eps_-$.
Then  by combining these contributions to the lepton asymmetry, 
we obtain:
\be
\epsilon =\epsilon_- + \epsilon_+ = \kappa (c_{\tilde L}-c_L) ,
\ee
where 
\be
\kappa =\left( \frac{16 B M_T ~{\rm Im}\Pi_{+-} }
{4 B^2 M^2_T +\Pi^2_{++}} \right)
\frac{\Gamma^L_{0}}{\Gamma} .
\label{epsilonT}
\ee
Notice that a non-zero asymmetry is obtained
because two important ingredients are present:
1) there are physical (unremovable) 
relative phases among the superpotential and related 
soft-trilinear couplings, ~2)
finite-temperature 
effects make the fermion and boson phase-spaces different, 
$c_L \neq c_{\tl}$.   
Another important feature of our leptogenesis scenario is that it
displays a resonance behaviour \cite{reson}
for $2 B M_T \simeq \Pi_{++}$ [see eq.~\rfn{epsilonT}], which depends on 
the temperature.

From $\eps$ the total produced baryon asymmetry $n_B/s$ 
 can now be calculated.
Since the asymmetry is temperature dependent, the  $n_B/s$ ratio 
cannot be expressed as the product of the CP-asymmetry times an 
efficiency factor which would be independent of this asymmetry.
As most of the terms in $\epsilon$ are 
temperature dependent,  all of the CP-asymmetry 
has to be absorbed in the efficiency factor which is the result of  
(numerical) temperature integration of the Boltzmann equations. 
Here, instead, to have some physical insight onto 
the various effects, we rewrite the asymmetry as
\be
\epsilon=\epsilon_0 \, C_T \, (c_{\tilde L}-c_L) ,
\label{epsdec}
\ee
with $\epsilon_0=\kappa |_{T=0}$ and $C_T=\kappa/(\kappa|_{T=0})$.
The three factors play  distinct roles: $i$) 
$\epsilon_0$ exhibits  the dependence of 
the asymmetry on 
the relevant interactions \rfn{mb} at zero temperature,  
$ii$) the factor $c_{\tilde L}-c_L$ 
encodes the crucial finite temperature effect leading to a non-vanishing 
asymmetry, $iii$) the coefficient $C_T$ reflects  residual (and minor) 
thermal effects and it is of order one.
In this way, the temperature-dependent factors 
$C_T$ and 
$c_{\tilde L}-c_L$ can be absorbed in the efficiency parameter $\eta$ 
and finally the baryon asymmetry can be expressed (at the leading-order 
in the soft terms) as:
\bea
\frac{n_B}{s}&=&-\left( \frac{24+4n_H}{66+13n_H}\right)
\epsilon_0
~\eta ~ 6 Y^{eq}_{T}
\nn\\
&=&-3\times 10^{-3} 
\epsilon_0 \eta .
\label{nBovers}
\eea
The first factor takes into account the reprocessing of the $B-L$
asymmetry by sphaleron transitions, with the number of Higgs doublets $n_H=2$.
The parameter
$Y^{eq}_{T}=45\zeta (3)/(2\pi^4 g_*)$ is
the equilibrium number density in units of entropy density\footnote{
In this supersymmetric extension of the standard model, the number 
of effective degrees of freedom is $g_* = 251.25$.} 
 for any 
component of the $T_-$, $T_-^*$, $T_+$ or $T_+^*$ triplets, 
and the factor 6 
accounts for the 6 states in $T_-$  and $ T_-^*$ (or for the 6  
states in $T_+$  
and $ T_+^*$).

\vspace{0.3cm}

\begin{figure}[t]
\centerline{\epsfxsize = 0.4\textwidth \epsffile{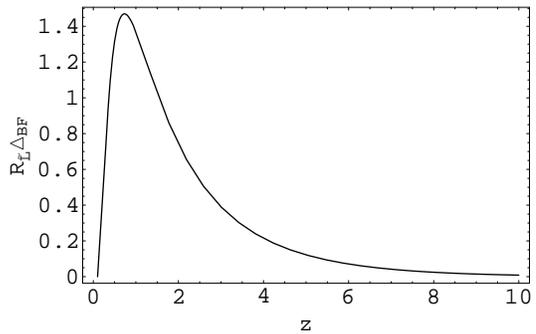}}
\caption{\it The quantity $R_{\tilde{L}} \Delta_{BF}$, defined 
in (\ref{xxx}), as a function of $z=M_T/T$.}
\label{cfcb}
\end{figure}

\mysection{{\bf 4.} Results.} 
For the sake of simplicity and discussion, let us 
make some reasonable assumptions to  show that our 
mechanism can lead to the observed baryon asymmetry. 
First, we neglect  the top Yukawa contribution\footnote{This may 
appear quite a questionable 
assumption, as at high energy-scale $10^{7}- 10^{10}~{\rm GeV}$ 
the coupling $y_t$ is generically comparable (or even larger) than 
the gauge couplings $g_2, g_1$. So neglecting 
the $y_t$ contribution is merely motivated by simplicity 
and do not spoil our illustrative presentation.}
in eq.~(\ref{Tmass}), so that 
all phase space  factors $c_a, c_{\tilde a}$, as well as $R_a, 
R_{\tilde a}$,  are 
equal for all scalar  states ${\tilde a}$  and for 
all fermion  states $a$.
Furthermore, we  also neglect the temperature effect in 
$\Pi_{++}$, as the maximum value the asymmetry can take 
(at the resonance) is hardly affected by this approximation. 
In addition, by assuming  `universal-like' 
relations for the trilinear couplings in eq.~\rfn{Lsoft}, {\it i.e.} 
$A_T =A Y_T, A_1= A \la_1, A_2= A \la_2$, 
the lepton asymmetry $\eps$ 
takes the simple form given by eq.~(\ref{epsdec}) with
\bea
&& \epsilon_0 \!= \!\left(\frac{16 \Gamma_0 B}{4 B^2 +\Gamma^2_0} 
\right)
\frac{{\rm Im} A}{M_T} \frac{\Gamma^L_{0}}{\Gamma_0} ,
\label{eps0}\\  
&&C_T\,(c_{\tilde L}-c_L) \approx 2 R_{\tilde L} \Delta_{BF}\,, 
\label{xxx}
\eea
where $\Gamma_0 = \sum_p \Gamma^p_{0}$
is the total decay width $(p=L,\tilde{L}$, $H_1, \tH_1, H_2, \tH_2)$ 
at zero-temperature 
[see eqs.~(\ref{gpart}),(\ref{gpart1})] and \cite{dgr} 
\be
\Delta_{BF}=\frac{c_{\tilde L}-c_L}{c_{\tilde L}+c_L} . \label{DFB}
\ee
In Fig.~\ref{cfcb} the temperature dependent factor  
$R_{\tilde{L}} \Delta_{BF}$ is plotted as a function of $z=M_T/T$.
For $T\gg M_T$ this factor and hence the asymmetry  
vanish  because no scalar decay modes
are kinematically available. For $T\sim M_T$ both the decays 
$T_\pm\to \bar{L} \bar{L}, {\tilde L}^* {\tilde L}^*$ 
become possible and so $\epsilon$ is unsuppressed 
due to the finite-temperature  effects. 
For $T\ll M_T$,  $\epsilon$ vanishes 
because the asymmetries in $L$ and $\tilde L$ exactly cancel 
against each other ($\Delta_{BF} =0$).
The asymmetry attains its resonant maximum value 
for $2 B= \Gamma_0$:  
\beq
\label{epsmx}
\epsilon^{max} =
8\frac{{\rm Im}A}{M_T} \frac{\Gamma^L_{0}}{\Gamma_0}
R_{\tilde L}\Delta_{BF} .
\eeq 
If, for example, we take  $M_T=10^8$ GeV, ${\rm Im}A\sim 1$ TeV 
and   $\Gamma^L_{0}\approx \Gamma$, from 
 eq.~(\ref{nBovers}) we see that   
the efficiency parameter needs to be 
$\sim {\cal O}(10^{-3})$ in order  to explain the observed  
$n_B/s\approx 8\times 10^{-11}$.
The efficiency $\eta$ is a function of $\Gamma$ and $M_T$ and, since 
the triplets have gauge couplings, 
$\eta$ is expected to be smaller
than in the singlet leptogenesis 
case~\cite{hms,tfermion}. Although we have not
solved the Boltzmann equations, 
we believe that,  
based on the results obtained in~\cite{hms,tfermion},  
$\eta\sim {\cal O}(10^{-3})$ is realistic.
%\footnote{
%A complete analysis to account, for example,  for a very 
%accurate calculation of the efficiency 
%parameter $\eta$ is beyond the scope of this letter.}.
For this estimate we use the efficiency factor 
given in Fig.~6 of Ref.~\cite{tfermion} for the case of 
a fermion triplet, under the 
assumption that it does not differ much from the one in the case of a 
scalar triplet.\footnote{Gauge scatterings are expected to be 
essentially the same 
up to factors of order one. 
A more precise determination of the produced baryon asymmetry, 
including in particular
the calculation of the efficiency factor for a scalar triplet,  is 
 beyond the scope of this letter and is left
for a subsequent publication.}
This estimate takes also into account the results of Fig.~2 as well as 
the fact that $\Gamma_0/H(M_T)$ is bounded from below by the value 
$\sim 40$, and hence some washout (though moderate) 
is active due to the inverse decay processes.
This bound on $\Gamma_0/H(M_T)$ is derived by minimizing  
the total width $\Gamma_0$,  which can be more conveniently expressed   
in terms of the neutrino masses $m_\alpha, \alpha =1,2,3$     
[see also eq.~\rfn{mnu}], {\it i.e.}:  
\be
\Gamma_0=\frac{M_T}{16 \pi}~\left( \frac{M^2_T}{\lambda^2_{2} v^4_2}~
 (\sum_\alpha m^2_\alpha) +\lambda_1^2+\lambda_2^2\right) .
\label{res}
\ee
The minimum is achieved for $\lambda_1=0$ and for 
\be
\lambda_{2} = \frac{ \sqrt{M_T \sqrt{\sum_\alpha m^2_\alpha}}}{v_2}\sim
4\times 10^{-4} \left( \frac{M_T}{10^{8}{\rm GeV}} 
\frac{m_3}{0.05{\rm eV}}\right)^{1/2} , 
\ee
where the neutrino masses are inferred by the 
present  data (and assuming hierarchical neutrino spectrum).
So the least value of $\Gamma_0$ is:
\be
\Gamma_{0}^{min}= \frac{1}{8 \pi} \frac{M_T^2 \sqrt{\sum m_\alpha^2}}{v_2^2} 
\sim 1~{\rm GeV} \left(\frac{M_T}{10^8~{\rm GeV}}\right)^2
\label{Gammamin}
\ee
and given that $H(T)\simeq 17~ T^2/M_{Pl}$, we get the aforesaid bound:  
%we see that the bound on $\Gamma_0/H(M_T)$ does not depend on $M_T$:
\be
\label{bound}
\frac{\Gamma_0}{H (T=M_T)} \gsim 40 \times 
\left(\frac{m_3}{0.05~{\rm eV}}\right) , 
\ee
which does not depend on $M_T$.
Now let us discuss the range of $M_T$ for which 
successful leptogenesis can be achieved.   
By considering eq.~\rfn{epsmx} and our subsequent discussion, 
we can see that leptogenesis  requires  
$M_T$ in the range $\sim 10^{3} -  10^{9}$~GeV
with the parameter ${\rm Im}(A)$ in the range $(0.1-10)$~TeV.
Higher values for $M_T$  would lead to a smaller baryon 
asymmetry because of the $1/M_T$ suppression in eq.~(\ref{epsmx}). 
Another important issue is the  size of the $B$-parameter 
implied by the triplet leptogenesis.
Consider the most favourable  situation 
when the  resonance is realized and taking the minimum total decay 
width of eq.~(\ref{Gammamin}),
{\it i.e.} with $B=\frac{1}{2}\Gamma_{0}^{min}$. Then 
for $M_T=10^8$~GeV ($10^9$~GeV) we have $B \sim 1$~GeV ($100~{\rm GeV}$).
Note that a value of $M_T$ as low as $1~{\rm  TeV}$,  
which is not prevented by the efficiency 
factor as shown in Ref.~\cite{tfermion},  requires 
a very tiny value of $B$,  of order $10^{-10}$~GeV,  which would 
call for  an explanation \cite{MW}.

\vglue 0.3cm

\mysection{{\bf 5.}~General remarks.}~For the 
sake of completeness,  we would like to comment on some additional 
related issues. First, in the 
supersymmetric case with gravitino mass 
close to the electroweak scale, avoiding the gravitino-overclosure 
of the Universe implies  
an upper bound $T_{RH}<10^8$ GeV on the reheating temperature~\cite{gr}. 
This is a potential problem in the singlet seesaw  leptogenesis 
 models which (assuming  hierarchical singlet neutrinos) 
require $M_{N_1}>2.4\times 10^9$ GeV \cite{di,gnrrs}
(see, however, ref.\cite{tfermion}).
This bound is  not in conflict with our triplet leptogenesis 
mechanism which can accommodate values of $M_T$ below $10^8$ GeV.

Another related issue concerns  the 
presence of such $SU(2)_L$ triplets at intermediate energy, 
below 
the grand unification scale $M_G$, which would spoil the gauge coupling 
unification. The latter can be maintained by embedding our scenario 
in a larger gauge theory,  such as, for example,  the $SU(5)$ model. 
In this case the triplets $T$ ($\bT$) fit into the $15$ ($\overline{15}$) 
representation, together with extra states $S$ (${\bar S}$) and 
$Z$  (${\bar Z}$) which decompose according to $SU(3)_c\times 
SU(2)_L\times U)1)_Y$ as 
$S \sim (6, 1, -2/3), Z \sim (3, 2,1/6)$. 
In the $SU(5)$ broken phase, we have 
the  couplings involving 
 $S$ and $Z$, namely $Y_S S d^c d^c, ~ Y_Z Z d^c L$ 
($d^c$ are the $SU(2)_L$ singlet down quarks, see \cite{anna} 
for more details), and those in 
eq.~\rfn{W} involving  the triplets $T, \bT$.
The interactions with the $SU(3)_c$-coloured Higgs partners   
$t_1$ and $t_2$, namely $S t_1 t_1, Zt_1 H_1 ({\bar S} t_2 t_2 ,
{\bar Z} t_2 H_2)$, do not appear as  $t_1$ and $t_2$ 
are decoupled at $M_G$ to prevent too rapid a proton decay.
Therefore, the interactions of $S$ and $Z$ with 
the matter fields below $M_G$ are $B-L$ conserving 
(while those of the triplets $T$ are not) and so 
they cannot play any role in the generation of the baryon asymmetry.
Though at $M_G$ the above interactions with the coloured Higgs 
states $t_1, t_2$ violate $B-L$, 
it is likely that any excess of 
$B-L$ number produced by these interactions is  washed-out 
by the $B-L$ violating interactions of 
the triplets $T, \bT$. The final produced asymmetry      
is therefore the one given only by the decays of the $SU(2)_L$ triplets.

The main attractiveness 
of the minimal supersymmetric type-II scenario 
is that the high-energy flavour structure can be 
univocally determined from the low-energy neutrino mass matrix.
We also  recall another appealing feature 
related to the above one. Namely,  
the Yukawa matrix $Y_T$ 
can induce radiative corrections  in the (soft-breaking) 
mass matrix $m_{\tl}$ of 
the sleptons $\tilde L$ and, as a result, lepton flavour violating 
 non-diagonal entries are generated, even if, for example,  $m_{\tl}$ is 
flavour-blind at $M_G$. 
Due to  the direct connection 
between $Y_T$ and the neutrino mass matrix, the 
flavour structure of the mass matrix $m_{\tl}$ can be determined 
in terms of the neutrino parameters  and 
so correlations among 
lepton flavour violating processes involving different 
families can be predicted \cite{anna}.
The leptogenesis scenario we have presented  depends not only 
on  superpotential couplings as the neutrino masses do, but also 
on (several) soft-breaking parameters, 
whose determination would require (and certainly deserve) 
 a proper phenomenological 
study to point up some correlations with other observables.
Testing our leptogenesis scenario appears to be as involved as 
 other high-scale realizations of leptogenesis.

\vspace{0.3cm}

\mysection{{\bf 6.} Conclusions}
In this paper we have 
shown that the minimal supersymmetric seesaw scenario 
with a single pair of triplets can generate successful leptogenesis 
through the $L$-violating decays of these states.
The necessary CP-asymmetry  arises from  the interference 
between supersymmetric and soft-breaking terms and 
by a resonant-enhancement due to the `soft' mass splitting 
of the triplets $T$ and $\bT$.  
In this scenario the mass of the triplet states 
should be in the range $(10^3 - 10^{9})~{\rm GeV}$
depending on the size of the soft-supersymmetry breaking 
parameters.

\vskip0.3cm
\noindent {\bf Acknowledgment.}
We thank M.~Pietroni, A.~Strumia, 
S.~Vempati and J. Valle for useful discussions.
The work of G.~D.~was partially supported by IHP-RTN,
EC contract No.\ HPRN-CT-2002-00311 (EURIDICE), 
of T.~H.~by the European Union (EU) 
Marie Curie HPMF-CT-01765 
and  HPRN-CT-2000-00152 (Susy in the Early Universe) 
contracts,   
of M.~R.~by the ESF Grants 5135 and 5935, by the
EC MC contract MERG-CT-2003-503626, and by the Ministry of Education and 
Research of the Republic of Estonia, of A.~R.~by 
the EU HPRN-CT-2000-00148 (Across the Energy Frontier)
and HPRN-CT-2000-00149 (Collider Physics) contracts.

\end{document}